\documentclass[journal]{IEEEtran}
\usepackage{amsmath,amssymb, amscd}
\usepackage{graphicx,dblfloatfix}
\usepackage[belowskip=-7pt,aboveskip=7pt]{caption}
\usepackage{multirow}
\usepackage{color}

\begin{document}
%
\title{Minimum  Distortion Variance Concatenated \\ Block Codes
for Embedded Source Transmission}

\author{Suayb S. Arslan  \\ Advanced Development Laboratory \\ Quantum Corporation, Irvine, CA
92617
 \\ Email: \texttt{Suayb.Arslan@Quantum.com}
 \thanks{The author is with Quantum Corporation, 141 Innovation Drive, Irvine, CA, 92617, USA. e-mail: Suayb.Arslan@Quantum.com. When this work was done,
 The author was with University of California, San Diego, La Jolla, CA. This work was supported by
LG Electronics Inc., the Center for Wireless Communications (CWC) at
UCSD, and the UC Discovery Grant program of the state of California.}} 

\maketitle

\begin{abstract}
Some state-of-art multimedia source encoders produce embedded source
bit streams that upon the reliable reception of only a fraction of
the total bit stream, the decoder is able reconstruct the source up
to a basic quality. Reliable reception of later source bits
gradually improve the reconstruction quality. Examples include
scalable extensions of H.264/AVC and progressive image coders such
as JPEG2000. To provide an efficient protection for embedded source
bit streams, a concatenated block coding scheme using a minimum mean
distortion criterion was considered in the past. Although, the
original design was shown to achieve better mean distortion
characteristics than previous studies, the proposed coding structure
was leading to dramatic quality fluctuations. In this paper, a
modification of the original design is first presented and then the
second order statistics of the distortion is taken into account in
the optimization. More specifically, an extension scheme is proposed
using a minimum distortion variance optimization criterion. This
robust system design is tested for an image transmission scenario.
Numerical results show that the proposed extension achieves
significantly lower variance than the original design, while showing
similar mean distortion performance using both convolutional codes
and low density parity check codes.
\end{abstract}

\section{Introduction}


Multimedia transmission for heterogeneous receivers is a challenging
problem due to the unpredictable nature of the communication
channels. Recent advances in  multimedia compression technology are
to account for an adaptation for the time-varying and band limited
nature of wireless channels. Progressive source coding is an
attractive solution for the transmission problems posed by
 multimedia streaming over such channels. The bit stream is generally said to be \emph{embedded} if the removal of the end parts of the source bit
 stream enables adaptations to end user preferences according to varying terminal and
 network conditions.
 For example, the
scalable extension of H.264 AVC \cite{SVC} allows reconstruction of
the video at various bit rates using partial bit streams
(\emph{layer}s) at the expense of some loss of coding efficiency
compared to the single layer counterpart \cite{H264}. Also, the  bit
streams produced by SPIHT \cite{SPIHT}, JPEG2000 \cite{JPEG2000} or
the MPEG-4 fine grain scalable (FGS) coding \cite{MPEG4} standards
are embedded and provide a bit-wise fine granularity in which the
bit stream can be truncated at any point for source decoding.
However, embedded source coders provide progressiveness at the
expense of possessing some features that make them vulnerable to
channel bit errors. For example, it is common to these source coders
that the usefulness of correctly received bits depends on the
reliable reception of the previous bits.  Therefore, an efficient
unequal error protection (UEP) scheme is needed for the reliable
transmission of such multimedia data. Conventionally, less
redundancy is added for each layer with decreasing importance for
decoding to allow a graceful degradation of the source at the
receiver \cite{Mohr}.

\begin{figure*}
\centering
\includegraphics[height=68mm, width=147mm]{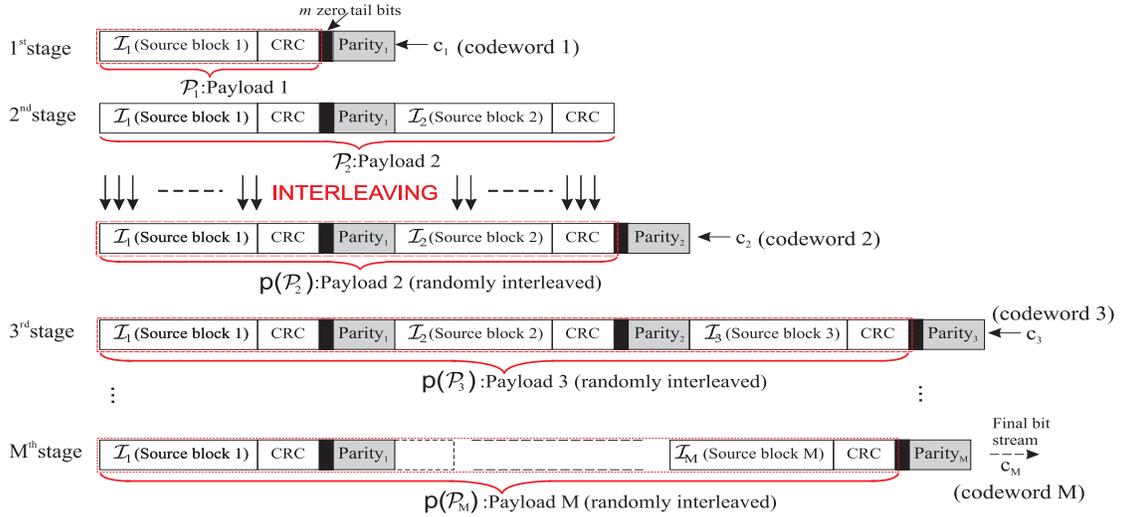}
\caption{\footnotesize{Proposed formatting consists of $M$ stages of
encoding as shown above.}} \label{encoding}
\end{figure*}

Transmission of progressive sources over error prone wireless
channels is a well investigated topic. Studies include various
cross-layer protection strategies for multimedia streaming over
wireless lossy networks \cite{Van} and adaptive selections of
application layer forward error correction (FEC) coding and
deployment for embedded bit streams \cite{Pan}, \cite{Stoufs}. For
the latter, joint source--channel coding (JSCC) is the most popular.
JSCC is extensively used  in the literature, in which an appropriate
channel code is used to protect the bit stream to optimize some
criterion such as minimization of mean distortion or maximization of
average useful source rate \cite{Cao}.

In a broadcast transmission scenario, each member of the network is
expected to receive at least a decent average multimedia quality in
order to meet the fair service guarantee. Excessive quality
fluctuations among the users of the same network can be avoided by
minimizing the variance of the distortion at the terminal of each
user \cite{Tagliasacchi}. The main contribution of this study is to
consider an efficient coding scheme in a broadcast scenario and
introduce major modifications to the original design of \cite{Suayb}
for improved distortion variance characteristics.

The concatenated block coded embedded bit streams are shown to give
superior performance over conventional coding paradigms while
providing flexible and low complexity transmission features over
multi-hop networks \cite{Suayb}.  There are two assumptions about
the previous coding structure that will not fit in a broadcast
transmission scenario. First of all, in the original coding scheme,
some of the information block sizes (optimized for minimum
distortion) might be very large. Typically, the optimal number of
encoding stages ($M^*$) are reported  to be four or five for the bit
budget constraints and raw channel bit error rates considered. This
means that there are five or six reconstruction levels at the
receiver. This may  not be desirable, for example, from an image
transmission perspective, because the user will only be able to see
at most six different quality versions of the transmitted image with
possible quality variations in between. Furthermore, it often leads
to user dissatisfaction.

Alternatively, each information block in the transmission system can
be chopped into smaller chunks to allow a larger number of
reconstruction possibilities at the receiver. Due to the embedded
nature of the bitstream, this can provide two advantages: (1) one
can obtain better mean distortion characteristics and (2) having
more reconstruction levels leads to user satisfaction and increases
the overall service quality. In other words, the image quality is
not expected to vary dramatically because of the availability of
larger set of reconstruction levels at the receiver. However, having
larger number of chunks in the system means more redundancy
allocation
 for error detection. Given the available
bit budget constraint, this will eventually leave less room for
source and channel coding bits. Thus, the paper is intended to carry
out the optimization needed to resolve this trade-off.

Secondly, the original optimization criterion was to minimize the
average distortion of the reconstructed source. Although this
criterion could be sufficient in a point-to-point communication
scenario, it is rarely found in a broadcast transmission scenario.
In order to maintain a decent average source quality among the
network users, the second order statistics of the source distortion
has to be taken into account. A way to approach this problem is
presented in this paper; we consider the minimum distortion variance
problem subject to a predetermined average source quality. This way,
a reasonable mean source distortion can be obtained  while
guaranteeing  the minimum deviation from the mean performance.

The remainder of this paper is organized as follows: In Section II,
the background information about concatenated block codes for
embedded source bit streams is explained in detail.
 In Section III, the proposed extension framework is presented and associated optimization criteria as well as the optimization problems are introduced.
 Some of the numerical results are given in Section IV. Finally, a brief summary
and conclusions follow in Section V.

\begin{figure*}
\centering
\includegraphics[width=149mm, height=23mm]{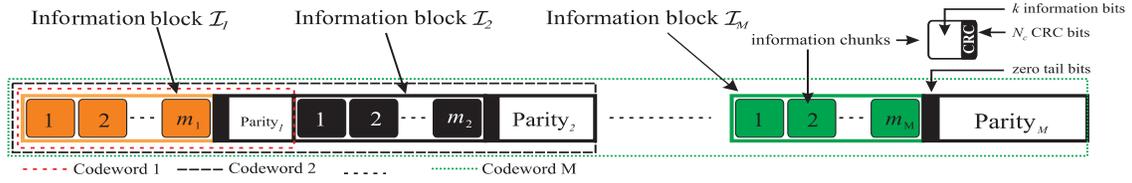}
\caption{\footnotesize{Encoding scheme using concatenated block
coding and chopped information blocks. $P_l$: parity bits for
codeword $l$. }} \label{Fig3}
\end{figure*}

\section{Concatenated block coding for embedded bit stream transmission}

Concatenated block codes are considered in \cite{Suayb} for embedded
bit stream transmission over error-prone memoryless channels. The
proposed $M$-codeword scheme is shown in Fig. \ref{encoding} and can
use any discrete code set $\mathcal{C}$. We give a brief description
of the original coding structure before giving the details of the
extension scheme.

We first describe the coding structure for convolutional codes. The
first stage of the encoder is the concatenation of $b_1$ source bits
(i.e., source block $\mathcal{I}_1$) with two bytes of
CRC\footnote{Here, a CRC polynomial is judiciously chosen to
minimize undetected-error probability and the same CRC polynomial is
used for all information blocks. The selected CRC polynomial is
$X^{16}+X^{12}+X^5+X$. Note that depending on the channel code used,
for example low density parity check codes, CRC bits may not even be
needed. }($N_c=16$ bits) based on $b_1$ bits for error detection. If
the convolutional codes are selected, they can still be treated as
block codes by appending  $m$ zero tailing bits to end the trellis
at the all zero state. Therefore,
 $|\mathcal{P}_1| = b_1 + N_c + m$ bits constitute  the first payload
$\mathcal{P}_1$. Later,  $\mathcal{P}_1$ is encoded using the
channel code rate $r_1 \in \mathcal{C}$ to produce the codeword
$c_1$. This ends the first stage of encoding. In the next stage,
$c_1$ is concatenated with the second information block
$\mathcal{I}_2$ (of size $|\mathcal{I}_2| = b_2$), $N_c$ and $m$
bits to produce the second payload $\mathcal{P}_2$ of size
$|\mathcal{P}_2| = \Bigl[(b_1 + N_c + m)/r_1\Bigl] + b_2 + N_c + m$.
In the next encoding stage, $N_c$ CRC bits are derived based only on
those $b_2$ bits. After the interleaving, the
 bits in $\boldsymbol{\pi}(\mathcal{P}_2)$   are encoded using code rate
$r_2 \in \mathcal{C}$ to produce codeword $c_2$ where
$\boldsymbol{\pi}(x)$ denotes the random block interleaving function
that chooses a permutation table randomly according to a uniform
distribution, and permutes the entries of $x$ bitwise based on this
table\footnote{We choose the size of the random permutation table to
be equal to the length of the payload size in each encoding stage
except the first.}. This recursive encoding process continues until
we encode the last codeword $c_M$. Lastly, the codeword $c_M$ is
transmitted over the binary symmetric channel (BSC) channel. Since
the errors out of a maximum likelihood sequence estimator (MLSE) are
generally bursty, and some of the block codes show poor performance
when channel errors are not independent and identically distributed
(i.i.d.) \cite{Forney1}, random block interleavers are used to break
up the long burst noise sequences.




The decoder performs the sequential encoding operations of the
encoder in reverse order on the noisy version ($\widehat{c}_M$) of
the codeword $c_M$. In other words, the noisy codeword
$\widehat{c}_M$ is decoded first using the corresponding channel
decoder and then, the deinterleaver is invoked to obtain
$\boldsymbol{\pi}^{-1}(\boldsymbol{\pi}(\mathcal{\widehat{P}}_M))=\mathcal{\widehat{P}}_M$.
Based on the decision of  the error detection mechanism (e.g. CRC
code), the $M$th information block ($\mathcal{I}_M$) is  labeled as
useful or not for the reconstruction process. Thus, $\mathcal{I}_M$
is associated with a label and peeled off from
$\mathcal{\widehat{P}}_M$. In the subsequent decoding stage,
$\widehat{c}_{M-1}$ is decoded and deinterleaved in the same manner
to obtain $\mathcal{\widehat{P}}_{M-1}$ and $\mathcal{I}_{M-1}$ is
determined to be useful or not in the reconstruction. The decoding
operation is finalized as soon as the decoding of codeword $c_1$ and
the label assignment of $\mathcal{I}_1$ are performed. Assuming that
the first label with a check failure is associated with
$\mathcal{I}_l$, then only the information blocks up to but not
including the block $l$ are used to reconstruct the source.

If we use low density parity check (LDPC) block codes, we do not
need to use $N_c$ CRC and $m$ tailing bits as the parity check
matrix of the code provides an inherent error detection capability.
However, similar to \cite{Pan}, an extra byte might be added for
each chunk to inform the RC-LDPC decoder about the channel coding
rate used for the next chunk. This can be thought of protocol based
redundancy allowed in the system and constrains the available bit
budget in transmission.

\section{Extension System and Optimization}

\subsection{Extension System}

In the original concatenated block coding scheme shown in Fig.
\ref{encoding}, there are $M$ encoding stages that produce a
sequence of embedded codewords. The number of reconstruction levels
at the receiver is $M+1$. As mentioned previously in the paper,
small $M^*$ leads to large variations in the quality of the
reconstructed source. In the extension system, each information
block $\mathcal{I}_l$ plus the corresponding $N_r$ redundant bits
(for example using convolutional codes we have $N_r = N_c$) are
chopped into smaller chunks of equal size ($\upsilon$ bits each) in
order to increase the number of reconstruction levels at the
receiver. This is illustrated in Fig. \ref{Fig3}. Each block of
($b_i + N_r$)-bits (we refer to this entity ``packet" later in the
paper) is constrained to be an integer multiple of $\upsilon$ bits
and the size of each information chunk is $k = \upsilon-N_r$ bits.
Let $m_l$ denote the number of separate chunks in the $l$th encoding
stage that makes up the block $I_l$ plus $N_c$ CRC bits in the
original encoding scheme for convolutional codes. Therefore, in the
extension scheme, we have $\sum_{l=1}^M m_l + 1 =\sum_{l=1}^M
\lfloor \frac{b_l+N_r}{\upsilon} \rfloor + 1$ number of
reconstruction levels. However, the proposed extension comes with
the cost that increasing the number of chunks increases the amount
of redundancy in the system. In the original design, total number of
source bits are $\sum_{l=1}^M \mathcal{I}_l$. In the proposed
extension however, since $\sum_{l=1}^M (m_l - 1) N_r$ extra
redundant bits are used, the number of source bits are given by
\begin{align}
\sum_{l=1}^M \mathcal{I}_l - (m_l - 1) N_r \leq \sum_{l=1}^M
\mathcal{I}_l.
\end{align}

\subsection{Minimum mean distortion and minimum distortion variance rate allocations}

In the original study, minimum mean distortion design criterion is
assumed \cite{Suayb}. Alternatively, we can minimize the
 distortion variance subject to a constraint on the mean
distortion performance. In other words,  the distortion variance can
be minimized such that the average distortion of the system is lower
than or equal to some predetermined mean distortion value
$\gamma_D$.

Let us assume that we are able to collect $r_{tr}$ channel bits per
source sample (e.g. pixels). We denote the available code rate set
by $\mathcal{C}=\{r_1,r_2,\dots,r_J\}$. Let us have $M$ encoding
stages having $m_i, 1\leq i \leq M$ chunks in the $i$-th coding
stage. For $i>M$, we define $m_{i}\triangleq1$ for completeness. We
use concatenated block coding mechanism to encode the information
chunks to produce the coded bit stream.  A \emph{code allocation
policy} $\pi$ allocates the channel code $c_{\pi}^{(i)} \in
\mathcal{C}$ to be used in the $i$-th stage of the algorithm. Note
that the number of packets in each information block depends on the
$\pi$ and therefore denoted as $m_i(\pi)$ hereafter. The size of the
outermost codeword length is given by
\begin{align}
r_{tr}  N_s  & =  \Biggl(\dots \biggl( \Bigl(\frac{m_1(\pi)\upsilon}{c_\pi^{(1)}}+m_2(\pi)\upsilon\Bigl) \frac{1}{c_\pi^{(2)}}+ \dots \biggl) \dots\Biggl)\frac{1}{c_\pi^{(M)}} \nonumber \\
& =  \sum_{i=1}^{M} \left( \prod_{j=i}^{M} c_{\pi}^{(j)}
 \right)^{-1} m_i(\pi) \upsilon
\end{align}
where $N_s$ is the number of source samples.

\emph{Assumption 1:} For a tractable analysis, we assume perfect
error detection.

\begin{figure*}[t!]
\begin{eqnarray}
 \overline{D}_{\pi}(n) = \sum_{j=1}^{M+1}  \sum_{i=0}^{ m_j(\pi)-1}
 D^n \left(
\bigg(\sum^{{j-1}}_{t=1} m_t(\pi) + i\bigg)\frac{k}{N_s}\right)
P_e(c^{(j, M)}_{\pi}) \left(1 - P_e(c^{(j, M)}_{\pi})\right)^i
\prod_{s=1}^{j-1} \left(1 - P_e(c^{(s, M)}_{\pi})\right)^{m_s(\pi)}
\label{Lemma1}
\end{eqnarray}
\hrule
\end{figure*}

For a given channel, let the probability of  decoding failure (for
example, CRC code flags a failure) for the chunk $i$ of the
information block $z$ (where $\sum_{j=1}^{z-1} m_j(\pi) < i \leq
\sum_{j=1}^z m_j(\pi)$), which is protected by the sequence of
channel codes $c^{(z)}_{\pi}, c^{(z+1)}_{\pi}, \dots, c^{(M)}_{\pi}
\in \mathcal{C}$, be $P_e(c^{(z, M)}_{\pi})$ for $z \leq M$. For
$z>M$, we define $P_e(c^{(z, M)}_{\pi})\triangleq1$. Let the
operational rate-distortion function of the source encoder be $D(R)$
where $R$ is the source rate in bits per source sample.

\emph{Assumption 2:} For the algorithm design purposes, we use a
similar approximation in \cite{Suayb} that decoder failure rate is
independent for each coded information chunk. This approximation is
shown to be good when convolutional and LDPC codes are used with
long enough interleavers \cite{Suayb}. In general, our code set
$\mathcal{C}$ can be chosen from any code family with a bit
processing method (such as interleaving) as long as this assumption
closely approximates the code block error performance.


\emph{Lemma 1:} Using \emph{Assumption 2}, $n$-th moment of  the
 distortion at the receiver
using the policy $\pi$,  $\overline{D}_{\pi}(n)$ is given by
Equation (\ref{Lemma1}).

\emph{Proof:} Let $X$ be a random variable that takes on the
distortion level $d$ with probability $p_d \triangleq Pr(X=d)$.
Consider the probability of truncating the chunk stream after
reliably receiving the $i$th chunk of the $j$th information block.
This corresponds to the source decoder that reconstructs the source
up to a distortion level $d_{j,i} = D\left( (\sum^{{j-1}}_{t=1}
m_t(\pi) + i)\frac{k}{N_s}\right)$, while the number of correctly
decoded chunks is $\sum^{{j-1}}_{t=1} m_t(\pi) + i$. Therefore,
\begin{align}
& p_{d_{j,i}} \triangleq P\left(X=D\bigg( (\sum^{{j-1}}_{t=1}
m_t(\pi) +
i)\frac{k}{N_s}\bigg)\right) \nonumber \\
& \ \ = P_e(c^{(j, M)}_{\pi}) \left(1 - P_e(c^{(j,
M)}_{\pi})\right)^i \prod_{s=1}^{j-1} \left(1 - P_e(c^{(s,
M)}_{\pi})\right)^{m_s(\pi)} \nonumber
\end{align}

Thus using \emph{Assumption 2}, the $n$-th moment of  distortion is
simply given as follows,
\begin{align}
&  \overline{D}_{\pi}(n) = \mathbb{E}[X^n] = \sum_{i,j} d_{j,i}^n
\times  p_{d_{j,i}} \nonumber
\end{align}

Finally, note that $j=1,\dots, M$ and $i=0,\dots,m_{j-1}(\pi)$
covers all the possibilities except the event that we receive all
the chunks correct. This is fixed by letting $j=M+1$ and
$m_{M+1}(\pi)=1$. $\blacksquare$

\subsection{Optimization Problems}

Next, we present the optimization problems considered in this study.
We start with the original optimization problem i.e.,
\emph{Minimization of Mean Distortion}, then we give the
\emph{Constrained Minimization of Distortion Variance} problem for
the proposed extension. Finally, we consider \emph{Minimum Second
Moment of Distortion} as an alternative solution for the latter.

\emph{Problem 1: (Minimization of Mean Distortion)}
\begin{align}
\min_{\pi,  \xi, \upsilon}  \overline{D}_{\pi}(1) \textrm{ such that
} r_{tr}
 = \frac{1}{N_s} \sum_{i=1}^{M}
  \frac{m_i(\pi) \upsilon}{\prod_{j=i}^{M}
r_{\pi}^{(j)}} \leq B \label{p1}
\end{align}
where $\xi=\{b_1,\dots,b_M\}$ and $B$ is some threshold transmission
rate in bits per source sample. As mentioned in the introduction
section, we are interested in the minimization of the distortion
variance subject to an average source quality constraint. This
problem can be formulated as follows

\emph{Problem 2: (Constrained Minimization of Distortion Variance)}
\begin{align}
\min_{\pi,  \xi, \upsilon} \sigma^2_{\pi}  \textrm{ such that }
r_{tr}
 = \frac{1}{N_s} \sum_{i=1}^{M}
  \frac{m_i(\pi) \upsilon}{\prod_{j=i}^{M}
r_{\pi}^{(j)}} \leq B, \overline{D}_{\pi}(1) \leq \gamma_D \nonumber
\end{align}
where
\begin{align}
  \sigma^2_{\pi} =  & \sum_{j=1}^{M+1}   \sum_{i=0}^{ m_j(\pi)-1} \left( d_{j,i} -
\overline{D}_{\pi} \right)^2 \times p_{d_{j,i}} =
\overline{D}_{\pi}(2) - \overline{D}_{\pi}^2(1) \label{E1}
\end{align}
and $\gamma_D$ is some mean distortion constraint on the average
performance of the extension system.

\emph{Problem 2} is relatively a harder problem than \emph{Problem
1} because now the each term of the sum in Equation (\ref{E1})
depends on the average distortion, which in turn depends on the
parameters of the system subject to optimization. This problem can
be simplified by the following observation.

Note that we have $ \overline{D}_{\pi}(2) \geq
\overline{D}_{\pi}^2(1)$ because by definition, the variance cannot
be negative. This means that the maximum value of
$\overline{D}_{\pi}(1)$ is upper bounded and when the equality holds
($\overline{D}_{\pi}(2) = \overline{D}_{\pi}^2(1)$), the variance is
minimized. On the contrary, if we allow lower
$\overline{D}_{\pi}(1)$ in order to obtain a better mean distortion,
we will get  a positive variance ($ \sigma^2_{\pi}
> 0$). Thus, it is reasonable to assume that $\sigma^2_{\pi}$ is a
non-increasing function of $\overline{D}_{\pi}(1)$  using the policy
$\pi$. In light of this assumption, we will set
$\overline{D}_{\pi}(1) =\gamma_D$ to end up with an easier problem
to solve:


\emph{Problem 3: (Minimization of the Second Moment of Distortion)}
\begin{eqnarray}
\min_{\pi} \overline{D}_{\pi}(2) \label{p3} \textrm{ subject to }
r_{tr}
 = \frac{1}{N_s} \sum_{i=1}^{M}
 \frac{ m_i(\pi) k }{\prod_{j=i}^{M}
r_{\pi}^{(j)}} \leq B
\end{eqnarray}

This problem gives the optimal solution of \emph{Problem 2} given
that it achieves the minimum when the mean distortion hits the
boundary of the constraint set. We solve aforementioned optimization
problems using numerical optimization tools. We  employ a
constrained exhaustive search to find the optimal code allocation
policy of the system.

\section{Numerical Results}

We  consider both the original as well as the extension schemes with
two different optimization criteria. In general, we have four
different possible combinations:

\begin{itemize}
\item \emph{ConMinAve}: Concatenated coding with minimum average distortion
optimization criterion. Let the minimum distortion be denoted as
$d^*$ at the optimum.
\item \emph{ConMinVar}: Concatenated coding with minimum distortion variance
optimization criterion.
\item \emph{ConChopMinAve}: Extension scheme with minimum average distortion
optimization criterion.
\item \emph{ConChopMinVar}: Extension scheme with minimum
distortion variance optimization criterion subject to a minimum
distortion constraint $\gamma_D \leq d^*$.
\end{itemize}

We do not consider the system ConMinVar, simply because we intend to
show how the ``chopping" method can be instrumental to improve the
performance of the original concatenated coding design. In addition,
 an increase in the distortion variance
performance is expected, as we allow worse mean average distortion
performance in the system.

Also, since we constrain the information packet size to be equal to
multiples of $\upsilon$ and that we have discrete number of code
rates in the code set $\mathcal{C}$, it is not always possible to
meet the average distortion constraint with equality i.e., $\gamma_D
= d^*$. Thus, in solving Problem 3, we allow a margin of $\zeta$ in
order to find the best approximate solution. In other words, in our
simulation results we have $|\gamma_D - d^*| < \zeta$.

\begin{figure}[t!]
\centering
\includegraphics[width=\columnwidth, height=71mm]{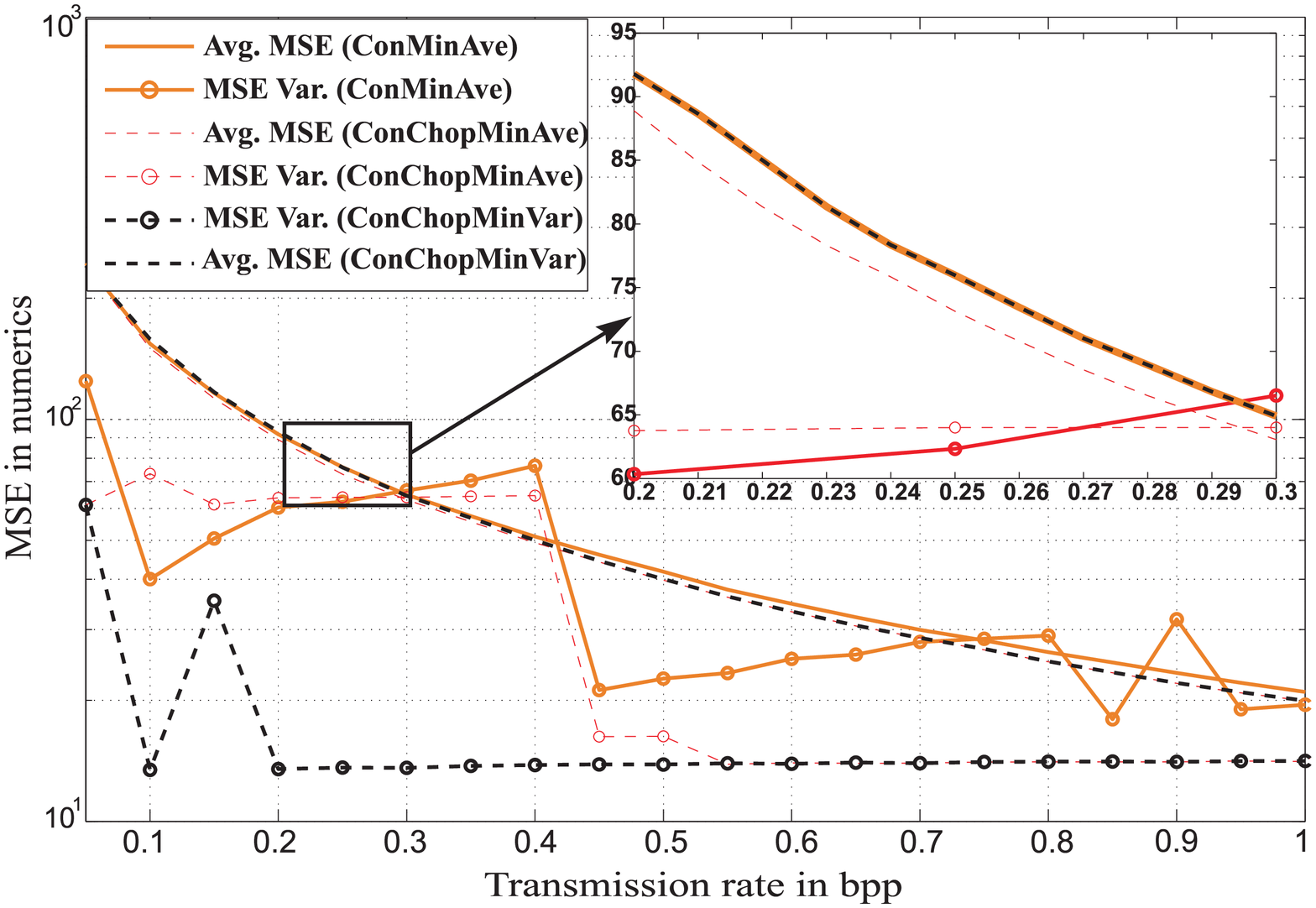}
\caption{{\footnotesize Comparisons of different systems under BSC
with crossover probability $\epsilon_0=0.05$ with $M=2$.}}
\label{meanvar1}
\end{figure}

We use $512 \times 512$ monochromatic images \emph{Lena} and
\emph{Goldhill} with SPIHT and JPEG2000 progressive image coders. In
the first simulation, we set $\upsilon=850$ bits, $M=2$ and use rate
compatible punctured convolutional (RCPC) codes with memory 6
\cite{Hagg}. We simulate all three systems and report average
distortion and distortion variance performances as functions of the
transmission rate in bits per pixel (bpp) when $\epsilon_0=0.05$. In
all the simulation results using RCPC codes, $\zeta \approx 0$ and
$\gamma_D \leq d^*$. As can be seen, chopping the information blocks
into smaller size chunks helps decrease the mean distortion and
distortion variance in almost all the transmission rates of
interest. In addition, allowing some performance degradation in mean
distortion, we can obtain much better distortion variance
characteristics.

\begin{figure}
\includegraphics[width=\columnwidth, height=70mm]{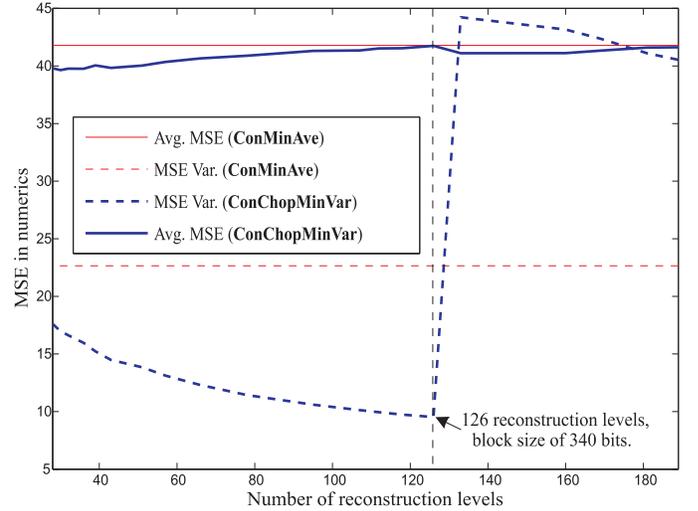}
\caption{{\footnotesize Comparisons of \emph{ConMinAve} and
\emph{ConChopMinVar} as a function of number of reconstruction
levels at the receiver under BSC with crossover probability
$\epsilon_0=0.05$ with $M=2$ and $r_{tr}=0.5$bpp.}} \label{meanvar2}
\end{figure}

\begin{table*}
\begin{center}
\begin{tabular}{|c|c|c|c|c|c|c|c|c|c} \noalign{\hrule height 1.5pt}
\multirow{2}{*}{Image} & \multirow{2}{*}{$r_{tr}$} &
\multirow{2}{*}{Results (Std. dev.)} & \multicolumn{3}{c|}{Channel
raw BER ($\epsilon_0$)} & \multirow{2}{*}{Image} &
\multicolumn{2}{c|}{Channel raw BER ($\epsilon_0$)}   \\
 \cline{4-6} \cline{8-9}
 &  &   & 0.1 & 0.05 & 0.01 &    & 0.05 & 0.01 \\  \noalign{\hrule height 1.5pt}
 \multirow{8}{*}{Lena} &  \multirow{3}{*}{0.25}&
ConMinAve & 89.9 & 62.33 & 52.68 &  \multirow{8}{*}{Goldhill}  &  73.95  & 59.95 \\
\cline{3-6} \cline{8-9}  &  &  ConChopMinVar & 19.62 & 12.79 & 8.58
& & 24.25
& 17.11 \\
\cline{3-6} \cline{8-9} &  & Percentage decrease & 78.17\% & 79.48\%
& 83.71\% & & 67.20\% & 71.46\%
\\ \cline{2-6} \cline{8-9}  &  \multirow{3}{*}{0.5} &
ConMinAve & 26.75 & 22.65 & 16.34 & & 99.92 & 18.73\\ \cline{3-6}
\cline{8-9} &
& ConChopMinVar & 16.33 & 9.53 & 7.66 & & 24.55 & 16.87 \\
\cline{3-6} \cline{8-9} (SPIHT) & & Percentage decrease & 38.95\% & 57.92\% & 53.12\% & (JPEG2000) & 75.43\% & 9.93\% \\
\cline{2-6} \cline{8-9}  & \multirow{3}{*}{0.8}
 & ConMinAve & 34.77 & 28.99 & 15.11 &  & 24.55 & 16.87 \\
\cline{3-6} \cline{8-9} & & ConChopMinVar & 16.03 & 4.92 & 2.65 & & 9.18 & 17.01 \\ \cline{3-6} \cline{8-9}  & & Percentage decrease & 53.9\% &  83.03\% & 82.46\% & & 73.93\% & 30.29\% \\
\noalign{\hrule height 1.5pt}
\end{tabular}
\end{center}
\caption{\footnotesize{Simulation results using RCPC codes. Average
performances are mean square error values in numerics.}}
\label{Table1}
\end{table*}

\begin{table*}
\begin{center}
\begin{tabular}{|c|c|c|c|c|c|c|c|c|} \noalign{\hrule height 1.5pt}
\multirow{2}{*}{Image} & \multirow{2}{*}{$r_{tr}$(bpp)} &
\multirow{2}{*}{Schemes } & \multirow{2}{*}{(Value)}  &
\multicolumn{2}{c|}{Channel raw BER ($\epsilon_0$)} &
\multirow{2}{*}{Image} &
\multicolumn{2}{c|}{Channel raw BER ($\epsilon_0$)}    \\
 \cline{5-6} \cline{8-9}
 &  & &  & 0.1 & 0.05 &     & 0.1 & 0.05  \\
\noalign{\hrule height 1.5pt}
 \multirow{10}{*}{Lena} &  \multirow{5}{*}{0.252}&
\multirow{2}{*}{ConMinAve} & Mean & 53.33(30.86) & 40.51(32.05) &  \multirow{10}{*}{Goldhill}  &  94.41(28.4)  & 79.38(29.13) \\
\cline{4-6} \cline{8-9} &  &  & Std. Dev. & \textcolor{blue}{19.7} &
\textcolor{blue}{47.25} & & \textcolor{blue}{22.4}
& \textcolor{blue}{6.53} \\
\cline{3-6} \cline{8-9} & &  \multirow{2}{*}{ConChopMinVar} & Mean &
54.03(30.8) & 40.81(32.02) & & 95.11(28.35) & 79.41(29.13) \\
\cline{4-6} \cline{8-9} &  &  & Std. Dev. & \textcolor{blue}{1.61} &
\textcolor{blue}{1.88} &  & \textcolor{blue}{1.1}
& \textcolor{blue}{1.62} \\
\cline{3-6} \cline{8-9} & & Percentage decrease & Std. Dev.&
\textcolor{red}{91.8}\%  & \textcolor{red}{96\%}  & &
\textcolor{red}{95\%} & \textcolor{red}{75.2\%}
\\ \cline{2-6} \cline{8-9} &  \multirow{5}{*}{0.505} &
\multirow{2}{*}{ConMinAve} & Mean & 25.29(34.08) & 18.96(35.35)  & & 59.27(30.41) & 46.05(31.5)\\
\cline{4-6} \cline{8-9} &  &  & Std. Dev. & \textcolor{blue}{0.41} &
\textcolor{blue}{2.34} &  & \textcolor{blue}{0.073}
& \textcolor{blue}{0.362} \\
\cline{3-6} \cline{8-9} &
& \multirow{2}{*}{ConChopMinVar} & Mean & 25.68(34.03) & 19.22(35.3)  & & 59.29(30.4) & 46.35(31.47) \\
\cline{4-6} \cline{8-9} &  &  & Std. Dev. & \textcolor{blue}{0.025}
& \textcolor{blue}{0.074} & & \textcolor{blue}{0.018}
& \textcolor{blue}{0.081} \\
\cline{3-6} \cline{8-9} & & Percentage decrease & Std. Dev. & \textcolor{red}{93.9\%} & \textcolor{red}{96.8\%}  & & \textcolor{red}{75.34\%} & \textcolor{red}{77.6\%} \\
\noalign{\hrule height 1.5pt}
\end{tabular}
\end{center}
\caption{\footnotesize{Simulation results using rate compatible LDPC
codes and JPEG2000 source coder. Average performances are mean
square error values in numerics. PSNR values in dB are included in
parentheses next to MSE results.}} \label{Table2}
\end{table*}

In the second simulation, we set $r_{tr}=0.5$bpp and $M=2$ and vary
$\upsilon$ to see the effect of variable chunk size on the overall
performance. First of all, smaller chunk size does not necessarily
mean better performance as the number of redundant CRC bits increase
and consume the available bit budget. Consider the system
ConChopMinVar. We have seen in the previous simulation that chopping
helps to improve the mean system performance. Thus, for a given $M$,
we can find an optimum chunk size that will minimize the distortion
variance given that it satisfies a mean distortion constraint. In
Fig. \ref{meanvar2}, we note that as we move from left to right on
the abscissa, the number of reconstruction levels increase i.e., the
block size decreases, number of blocks increases and number of
redundancy used for error detection in the bit budget increases.
Also, we observe that as we sacrifice some mean distortion
performance, we obtain a decrease in distortion variance. This
numerical example shows the validity of Assumption 2 about the
relationship between the mean distortion and the distortion
variance. They are observed to be inversely related.

In Fig. \ref{meanvar2},  we observe that  the minimum variance is
achieved when the block size hits 340 bits while satisfying the
desired mean distortion constraint $d^* = 41.79$ with equality. At
the optimum, ConMinAve has only 3 reconstruction levels (since
$M=2$) at the receiver while ConChopMinVar has 126 different
reconstruction levels. ConMinAve has a variance of $22.65$ and shown
as a horizontal line for comparison. The variance of ConChopMinVar
shows a jump after achieving the optimum at a variance of $9.53$
(almost $\%58$ percent decrease from that of ConMinAve). This is
because as we have more chunks and therefore more reconstruction
levels, CRC bits become dominant in the system. In order to satisfy
the mean distortion constraint, the optimization mechanism changes
the optimum channel code rates from $(4/5,4/9)$ to $(8/9,4/11)$.
Having more powerful protection now decreases the mean distortion
value while causing an increase in the total variance. Thus,
ConChopMinVar has $\%52$ less distortion variance compared to
ConMinAve while both systems have almost the same mean distortion
characteristics. Table \ref{Table1} presents a set of performance
results using different images, transmission rates at various raw
channel BERs. As can be observed, dramatic improvements on the
variance characteristics of the original design are possible using
the extension system.

Finally in Table \ref{Table2}, we provide some of the simulation
results using rate compatible LDPC codes \cite{Zaheer}.  We observe
that $\gamma_D \approx d^*$ (i.e., $\max \zeta = 0.7$) can be
achieved using LDPC codes. However, we can obtain dramatic
improvements in variance performance at the expense of little loss
in expected distortion performance of the original design. Table
\ref{Table1} presents a set of performance results using different
images, transmission rates at various raw channel BERs considered in
\cite{Pan} and \cite{Suayb}. As can be observed, similar performance
gains are possible. For example at a transmission rate
$r_{tr}=0.505$bpp and $\epsilon_0 = 0.05$, the ConMinAve chooses
$(4/5,2/3)$ as the two optimal code rates with three levels of
reconstruction since $M=2$. In the extension scheme ConChopMinVar,
choosing $\upsilon = 2000$bits and using the same optimal code rate
pair, we obtained 44 different levels of reconstruction. The latter
design gives almost the same image quality ($\sim 35.3$dB) with a
dramatic improvement in the variance, i.e., around $ 96.8\%$
decrease in variance compared to the that of ConMinAve.

\section{Conclusions}

We have considered minimum variance concatenated block encoding
scheme for progressive source transmissions. A non-trivial extension
of the original design is introduced with better reconstruction
properties at the receiver and more importantly better distortion
variance characteristics at a given average reconstruction quality.
We have considered three different optimization problems and
simplified the variance distortion minimization problem.  Simulation
results show that dramatic improvements can be obtained with the
extension system compared to the original coding scheme.


\begin{thebibliography}{1}


\bibitem{SVC}  T. Wiegand, G. Sullivan, J. Reichel, H. Schwarz, M. Wien (Editors),
``Joint draft 9 of SVC amendment (revision 2)," Document JVTV201,
Marrakech, Morocco, January 13-19, 2007

\bibitem{H264} ITU-T Rec. H.264 \& ISO/IEC 14496-10 AVC, ``Advanced Video
coding for generic audiovisual services," Version 3, 2005.

\bibitem{SPIHT} A. Said  and W. A. Pearlman, ``A New Fast and Efficient Image Codec Based on Set Partitioning in
Hierarchical Trees,"  \emph{IEEE Trans. on Circuits and Systems for
Video Tech.,} vol. 6, pp.243-250, June 1996.

\bibitem{JPEG2000} A. Skodras, C. Christopoulos, and T. Ebrahimi, ``The JPEG2000 still
image compression standard," \emph{IEEE Signal Process. Mag.}, vol.
18, no. 9, pp.36–58, Sep. 2001.

\bibitem{MPEG4} ``Streaming video profile- Final Draft Amendment
(FDAM 4)," ISO/lEC JTCl/SC29/WGI lN3904, Jan. 2001.


 \bibitem{Mohr}A. E. Mohr, E. A. Riskin, and R. E. Ladner, ``Unequal
loss protection: Graceful degredation of image quality over packet
erasure channels through forward error correction," \emph{IEEE J.
Sel. Areas Commun.,} vol. 18, no. 6, pp.819-860, Jun. 2000.

\bibitem{Van} Van der Schaar, M. \& Sai Shankar, D. ``Cross-layer wireless multimedia transmission:
Challenges, principles and new paradigms," \emph{IEEE Trans. on
wireless Communications,} Volume: 12, Issue: 4, pp.50-58, 2005

\bibitem{Pan} X. Pan, A. Cuhadar, and A. H. Banihashemi, ``Combined source
and channel coding with JPEG2000 and rate-compatible low-density
parity-check (RC-LDPC) codes," \emph{IEEE Trans. Signal Process.,}
vol. 54, no. 3, pp. 1160–1164, Mar. 2006

 \bibitem{Stoufs} M. Stoufs, A. Munteanu, J. Cornelis, and P. Schelkens, ``Scalable joint
source-channel coding for the scalable extension of H.264/AVC,"
\emph{IEEE Trans. Circuits Syst. Video Technol.,} vol. 18, no. 12,
pp. 1657--1670, Dec. 2008.

 \bibitem{Cao} L. Cao, ``On the unequal error protection for progressive image
transmission," \emph{IEEE Trans. on Image Processing,} vol.16, no.9,
Sept. 2007.


 \bibitem{Tagliasacchi} M. Tagliasacchi, G. Valenzise, and S. Tubaro, ``Minimum variance
optimal rate allocation for multiplexed H.264/AVC bitstreams,"
\emph{IEEE Trans on Image Processing,} vol. 17, no. 7,
pp.1129--1143, July 2008.


\bibitem{Suayb} S. S. Arslan, P. C. Cosman, and L. B. Milstein, ``Concatenated Block Codes for Unequal Error Protection of Embedded Bit Streams,"
\emph{IEEE Trans. on Image Processing}, vol. 21, no. 3, pp.
1111--1122, Mar. 2012.


\bibitem{Forney1} G. D. Forney,  ``Concatenated Codes,"  MIT Press, Cambridge, MA, 1966.

\bibitem{Hagg} J. Hagenauer, ``Rate-Compatible Punctured Convolutional Codes (RCPC Codes) and Their Applications," \emph{IEEE Trans. on
Commun.}, vol. 36, No. 4, pp.389--400, April 1997.

\bibitem{Zaheer} S. F. Zaheer, S. A. Zummo, M. A. Landolsi and M. A. Kousa, ``Improved regular
and semi-random rate-compatible Low-density parity-check with short
block lengths" , \emph{IET Commun.,} vol. 2, no. 7, pp. 960--971,
2008.








\end{thebibliography}
\end{document}